\documentclass[a4paper]{jpconf}
\usepackage{graphicx}
\usepackage{tabularx}
\usepackage{bm,color,amsmath,amssymb,mathrsfs,latexsym,graphicx,psfrag}

\begin{document}
%\nocite{*}

\title{Theory of `hidden' quasi-1D superconductivity in Sr$_2$RuO$_4$}

%\noindent \qquad \\[-6pt] \qquad Version 2.0\\\qquad December 21, 2006

\author{S Raghu, Suk Bum Chung and Samuel Lederer}

\address{Department of Physics, Stanford University, Stanford, California 94305, USA}

\ead{sraghu@stanford.edu}

\begin{abstract}
Is the mechanism of unconventional superconductivity in Sr$_2$RuO$_4$ closer in spirit to superfluid $^3$He, or to the cuprates, pnictides, and organic superconductors?    We challenge prevailing assumptions in this field and using well-controlled perturbative renormalization group calculations, we suggest that superconductivity in Sr$_2$RuO$_4$ resembles more closely the quasi-one dimensional organic superconductors.  Our theory has certain phenomenological consequences that are consistent with the experimentally observed phenomena.  
\end{abstract}

\section{Introduction}

Strontium Ruthenate, Sr$_2$RuO$_4$, 
is in many ways an archetypal unconventional superconductor.  
We say this for several reasons.  Firstly, Knight shift measurements\cite{Ishida1998} convincingly show that the ground state is a spin-triplet (i.e. odd-parity\cite{Nelson2004,Kidwingira2006} ) superconductor.  In a metal with time-reversal and inversion symmetry, such states cannot arise from the electron-phonon interactions and must occur as a consequence of  the bare {\it repulsive} electron-electron interactions.  Secondly,  its layered perovskite structure can be synthesized to a high degree of perfection, permitting us to neglect the effects of disorder.  Thirdly, its  electronic structure is relatively simple compared to related unconventional superconductors such as heavy fermion systems.  Most importantly, its normal state is a well-behaved Fermi liquid, as is known from  quantum oscillations\cite{Mackenzie1996, Bergemann2003}, heat capacity\cite{Deguchi2004}, and photoemission experiments\cite{Damascelli2000}(For a comprehensive review, see Ref. \cite{Mackenzie2003}).  Therefore, it should be possible to construct a microscopic theory of unconventional superconductivity that is based on well-controlled effective field theoretic descriptions of a Landau Fermi liquid theory and its instabilities.  Such a theory would be an anchor point for understanding other unconventional superconductors such as the organic, and perhaps even some aspects of the Iron pnictide and  cuprate superconductors.

This conference proceeding is based on the following works\cite{Raghu2010, Raghu2010a,Chung2012}.  However, in addition to elaborating and clarifying some aspects of this work,  we will introduce some new points of perspective that we hope may be of some use to those working in this fascinating field.  

\section{Electronic structure considerations}
\begin{figure}
\begin{center}
\includegraphics[width=5.0in]{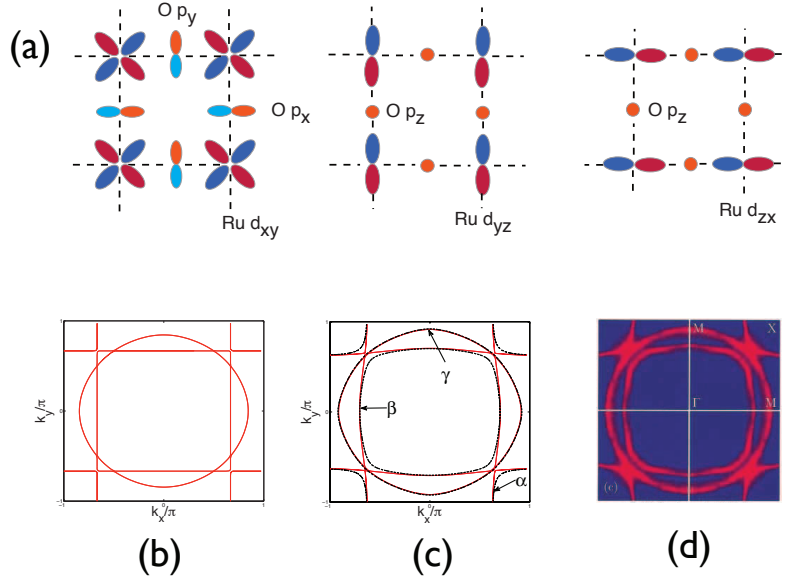}
\end{center}
\caption{Electronic structure of Sr$_2$RuO$_4$.  The Fermi surfaces are derived from the Ru t$_{2g}$ electrons.  Electrons in the d$_{xy}$ orbital tunnel equally well in both directions, and their dispersion produces the circular Fermi surface shown in the photoemission data (d).  By contrast, electrons in d$_{xz}$(d$_{yz}$) orbitals tunnel mainly along the x(y) direction.  Their quasi-one dimensional character produce the straight portions of the Fermi surface.  When orbital mixings are excluded, the Fermi surface of a simple tight-binding model has the form in (b); including the lowest order inter-orbital mixings produce the Fermi surface in (c), which is an excellent approximation to the experimentally observed Fermi surface (d).    }
\label{orbitals_fs}
\end{figure}
The quasiparticle bands at the Fermi energy of this system are derived primarily from the Ruthenium d$_{xz}$ d$_{yz}$, d$_{xy}$ orbitals.  The electrons in these orbitals tunnel to neighboring sites by making use of intervening oxygen atoms; however, for a low energy description that focuses on states near the Fermi level and preserves all the symmetries of the system, it is sufficient to consider solely the d-orbitals.  We discuss the electronic structure at a qualitative level in this section; further details will be found in Sec. \ref{weakcoupling}.  Due to the tetragonal symmetry of the system, the d$_{xz}$, d$_{yz}$ orbitals remain degenerate and transform as a doublet under point group operations whereas the d$_{xy}$ orbital is non-degenerate.  The electron kinetic energies are strongly dependent on the orbital character of the wave functions (Fig. \ref{orbitals_fs}).  Electrons in a d$_{xy}$ orbital tunnel with equal amplitude along the x,y nearest neighbor displacements.  This gives rise to a fairly isotropic Fermi surface, shown in Fig. \ref{orbitals_fs}.  By contrast, electrons in the d$_{xz}$ d$_{yz}$ orbitals have a quasi-one dimensional character: an electron in a d$_{xz}$(d$_{yx}$)  orbital has a substantially larger nearest-neighbor tunneling amplitude along the $\hat x$($\hat y$) direction.  This produces the highly anisotropic Fermi sheets in Fig. \ref{orbitals_fs}.  In the artificial limit where there is no mixing among the different orbitals, the Fermi surfaces are shown in Fig. \ref{orbitals_fs}b.  However, when spin-orbit coupling and longer range hoppings between the d$_{xz}$ , d$_{yz}$ orbitals are included, the Fermi surfaces take the form shown in Fig. \ref{orbitals_fs}c, which is an excellent approximation to the experimentally observed Fermi surface, reproduced from Ref.\cite{Damascelli2000,Veenstra2012} in Fig. \ref{orbitals_fs}d.  It is important to stress that the bandstructure here incorporates renormalization effects due to Fermi liquid parameters, such as mass enhancements, etc.  Below, when we present a weak-coupling theory, it is to be understood that while the bare electron-electron interactions are always large, the residual interactions between dressed quasiparticles are assumed to be weak compared to their kinetic energy.  

In the experimentally observed Fermi surface, there are three distinct sheets, denoted $\alpha,\beta, \gamma$.  The $\gamma$ sheet represents the circular Fermi surface and consists primarily of electrons in the d$_{xy}$ orbital.  The $\{ \alpha, \beta \}$ sheets are quasi-one dimensional and are primarily built from d$_{xz}$, d$_{yz}$ orbitals.  Due to the tetragonal symmetry and the distinct orbital character of the three Fermi surfaces, the superconducting pairing is believed to be predominantly derived from  either the $\gamma$  or the $\{ \alpha, \beta \}$ bands; a superconducting state that develops equally in all 3 sheets requires some fine-tuning and is therefore non-generic\cite{Agterberg1997}.    One of the open problems in the microscopic theory is to identify the so-called ``active" bands.  Below T$_c$, superconductivity will be induced in the remaining ``passive" degrees of freedom via a proximity effect.  Thus, while at zero temperature, the system behaves as a p-wave superconductor with gaps on all 3 Fermi sheets, there are a range of circumstances  in which the system behaves as though only the active bands exhibit appreciable pairing.  

\section{Orbital dependent superconductivity}
The prevailing assumption in the field has been that  $\gamma$ is the active band.  In addition to making our lives easier, there are good reasons for supposing this to be the case.  Firstly, from quantum oscillation studies, it is known that the  $\gamma$ band has the largest mass enhancement relative to the values based on first principles calculations.  Secondly, it is close to a van Hove singularity, which makes it natural to believe that ferromagnetic fluctuations could play a key role in mediating spin-triplet pairing among these electrons.  Such a scenario would be reminiscent of the physics of $^3$He, where ferromagnetic fluctuations (paramagnons) are thought to play a key role.  However, as can be seen from the experimental Fermi surface in Fig. \ref{orbitals_fs}(d), $\gamma$ is nearly perfectly circular; this makes the particle-hole susceptibility nearly a constant, as we discuss below, and paramagnon mediated spin-triplet pairing is much weaker in such a system.

	More importantly, inelastic neutron scattering studies have revealed that the in-plane magnetic excitations are predominantly  antiferromagnetic in character\cite{Sidis1999}; these are derived from the reasonably well-nested $\{\alpha, \beta\}$ Fermi surfaces.    If these degrees of freedom were the active bands, then spin-triplet superconductivity would arise mainly from incommensurate particle-hole fluctuations, and would be closer to what is found in the quasi-one dimensional organic superconductors.  This dichotomy, namely whether the microscopic pairing mechanism in Sr$_2$RuO$_4$ is closer to $^3$He or to the organics, cuprates and pnictides,  was emphasized in a recent review by D. Scalapino\cite{ScalapinoRMP}.  Despite the observation of incommensurate antiferromagnetic fluctuations and the absence of clear ferromagnetic correlations, the idea that $\gamma$ is the active band, and that superconductivity in this material is reminiscent of superfluid $^3$He, has persisted.  
\begin{figure}
\begin{center}
\includegraphics[width=5.0in]{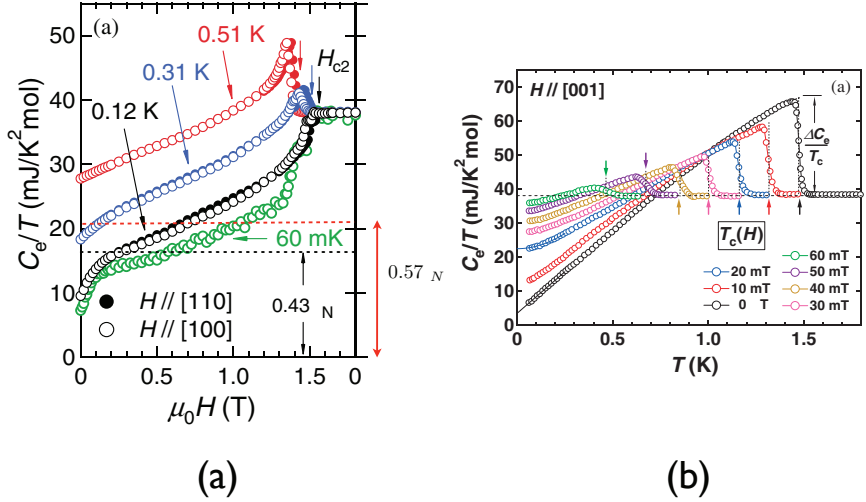}
\end{center}
\caption{Data from \cite{Deguchi2004}.  (a) Heat capacity in an in-plane magnetic field.  The black dashed line corresponds to $43 \%$ of the normal state density of states and falls in the ``shoulder" region.  We have added the red dashed line, which corresponds to $57 \%$ of the normal state density of states.  This line equally well falls into the shoulder region.  Both these lines are estimates of the remaining density of states of the passive bands assuming that a full gap opens on the active band(s).  The data is consistent with either scenario for the active bands.  (b) Heat capacity as a function of temperature.  Note that the ``two gap" feature is absent at zero field.   }
\label{cv}
\end{figure}
	
The main experimental evidence that is cited in favor of $\gamma$ as the active band is the heat capacity measurements of Ref.\cite{Deguchi2004}.  In this beautiful work, the specific heat was obtained in ultra-pure samples both as a function of temperature and pressure.  Since the specific heat coefficient is directly proportional to the total density of states at the Fermi level, $C/T \sim k_B^2 \rho(E_F)$, the authors, using knowledge from dHvA oscillations, attempted to identify the degrees of freedom that become gapped below the superconducting transition.  From the dHvA measurements, it is known that $57 \%$ of $\rho(E_F)$ is derived from the $\gamma$ sheet and $\{\alpha, \beta \}$ contribute to the remaining $43 \%$\cite{Mackenzie1996}.  Indeed, the heat capacity data as a function of field shows features suggestive of a two-gap superconductor (Fig. \ref{cv}a).  There is a sharp anomaly near the upper critical field, below which there is a ``shoulder" at lower fields, followed at even lower fields by another feature.   The authors of ref. \cite{Deguchi2004} claimed that the shoulder region matches well with the value of $ 43 \%$ of the normal-state density of states, from which they concluded that the $\{\alpha, \beta \}$ bands are the passive degrees of freedom.  

However, there are several reasons to challenge this interpretation.  Firstly, there is an underlying assumption that the pairing gap is fully isotropic on both the active and the passive bands so  a sharp separation may be invoked in identifying the source of the shoulder in the heat capacity.  	Secondly, there is a disagreement of approximately $20 \%$ between the heat-capacity measurements and the dHvA measurements for the total normal state density of states.  Thirdly, the ``shoulder" region is sufficiently wide that even the opposite scenario in which the $\{\alpha, \beta \}$ bands are the active ones, could account for it.  We have added a dashed red line in Fig. (\ref{cv}a) that would correspond to the case in which $43 \%$ of the normal-state specific heat derives from the active bands, which also lies in the shoulder region of the data.  Lastly, the two-gap features are not seen in the temperature dependence of the heat-capacity in the absence of a field, which is perplexing.  The data is undeniably important and possesses many clues as to the nature of the microscopic pairing structure.  However taken at face value, it does not seem to rule out either scenario for the active degrees of freedom in this system.

\section{Weak-Coupling theory}
\label{weakcoupling}
Having shown that the heat capacity measurements do {\it not}  unambiguously favor either scenario for the active band(s), we now turn to the main topic and present a microscopic theory of superconductivity.  We show that in the weak-coupling limit, there is a very sharp and inescapable conclusion: the active bands are $\{ \alpha, \beta\}$, and triplet pairing in this system is reminiscent more of the quasi-one dimensional organics than of superfluid $^3$He.  Using a simple extension of a recently 
 developed asymptotically exact weak-coupling analysis of the Hubbard model\cite{Raghu2010}, we show that the dominant superconducting instability is in 
 the triplet channel and occurs among the quasi-1D Fermi surfaces of Sr$_2$RuO$_4$.  
 
We consider the dynamics of electrons  in a tetragonal lattice of  Ru $t_{2g}$ orbitals governed by a Hamiltonian of the form 
\begin{equation}
H = H_0 + U \sum_{i \alpha} n_{i \alpha \uparrow} n_{i \alpha \downarrow} + \frac{V}{2} \sum_{i, \alpha \ne \beta} n_{i \alpha} n_{i \beta} + \delta H
\end{equation}
Here, $n_{i \alpha \sigma}$ is the density of electrons having spin $\sigma$ at position $i$ in orbital $\alpha = xy,yz,zx $, and $n_{i \alpha} = \sum_{\sigma} n_{i \alpha \sigma}$.  The strength of the repulsive interaction between two electrons on like (distinct) 
orbitals at the same
lattice site is given by U(V).    
$H_0 = \sum_{\alpha} \sum_{\vec{k} \sigma} \left(\epsilon^0_{\alpha \vec{k} } - \mu \right) c^{\dagger}_{\vec k \alpha \sigma} c_{\vec k \alpha \sigma} $ is the dominant intra-orbital kinetic energy and gives rise to three decoupled energy bands at the Fermi level as shown in Fig. \ref{orbitals_fs}.  Here, we make use of the following tight-binding parametrization of these energies\cite{Kontani2008,Liu2008}:
\begin{eqnarray}
\epsilon^0_{xz}(\vec k) &=& -2t \cos{k_x} - 2t^{\perp}\cos{k_y} \nonumber \\
\epsilon^0_{yz}(\vec k) &=& -2t^{\perp} \cos{k_x} - 2 t \cos{k_y} \nonumber \\
\epsilon^0_{xy}(\vec k) &=& -2t' \left( \cos{k_x} + \cos{k_y} \right) - 4 t'' \cos{k_x} \cos{k_y} 
\end{eqnarray}
with $(t,t^{\perp},t', t'', \mu) = (1.0,0.1,0.8,0.3, 1.0)$.  The quantity $\delta H$ represents 
smaller terms such as longer range hopping and spin orbit coupling which mix the distinct 
orbitals.   
It plays a relatively minor role in determining the superconducting transition temperature, and can therefore be considered perturbatively.  However, it plays a crucial role in selecting a superconducting state which breaks time-reversal symmetry, as we discuss below.  When $\delta H = 0$, the non-interacting  charge and spin susceptibilities of the normal state are separate functions for each orbital:
\begin{equation}
\chi_{\alpha}(\vec q) = -\int \frac{d^2 k}{\left( 2 \pi \right)^2} \frac{f(\epsilon_{\alpha, \vec k + \vec q}) - f(\epsilon_{\alpha, \vec k})}{\epsilon_{\alpha, \vec k + \vec q} - \epsilon_{\alpha, \vec k}}
\end{equation}
where $f(\epsilon)$ is the Fermi distrubition function, and the total susceptibility is obtained by summing the contributions from all three orbitals.  Since the quasi-two dimensional band is 
almost circular with a radius $k_f^{2d}$, its susceptibility is nearly  constant $\chi_{xy} \sim 1/4 \pi t'$ for $q < 2k_f^{2d}$, 
whereas the quasi-1D bands have a susceptibility that is peaked at a wave-vector $\vec q_x = (2k_f^{1d}, \pi)$ for the xz orbital and $\vec q_y = (\pi, 2k_f^{1d})$ for the yz orbital.  It is this latter feature which, in the weak-coupling framework, gives rise to the incommensurate spin fluctuations in the material\cite{Sidis1999}.  
\begin{figure}
\begin{center}
\includegraphics[width=3.in, angle=-90]{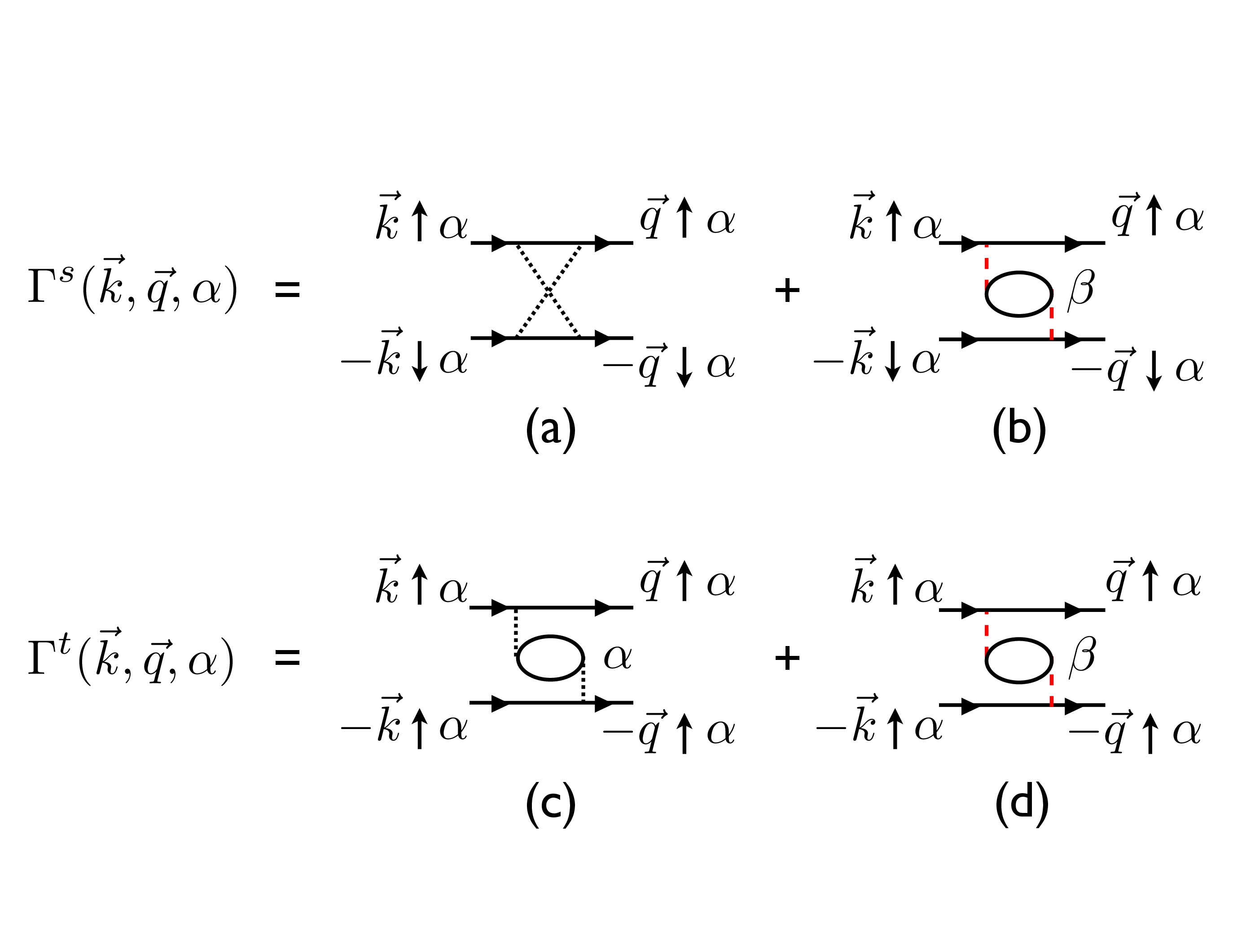}
\caption{Lowest order Feynman diagrams which contribute to the effective 
interaction between electrons within the same orbital $\alpha$ in the Cooper channel.  The dotted (black) line corresponds to the 
intra-orbital repulsion $U$ and the dashed (red) line corresponds to the inter-orbital repulsion V.       }
\label{diagrams}
\end{center}
\end{figure}

Since the superconductivity in Sr$_2$RuO$_4$ evolves out of a Fermi liquid with $T_c \ll E_f$, we proceed 
with a weak coupling analysis and consider the limit $U,V \ll  W$ where $W$ is the bandwidth.  
In this limit, superconductivity is the only instability of the Fermi liquid, and it can be treated in an asymptotically exact manner via a two-stage renormalization group analysis\cite{Raghu2010}.  In the first stage, high energy modes are perturbatively integrated out above an unphysical cutoff, and an effective particle-particle interaction in the Cooper channel is derived:
\begin{eqnarray}
\label{gamma}
\Gamma_s(\hat k, \hat q, \alpha) &=& U^2 \chi_{\alpha}(\hat k + \hat q) - 2V^2 \sum_{\beta \ne \alpha} \chi_{\beta} (\hat k - \hat q) \nonumber \\
\Gamma_t(\hat k, \hat q, \alpha) &=& -U^2 \chi_{\alpha}(\hat k - \hat q) - 2V^2 \sum_{\beta \ne \alpha} \chi_{\beta} (\hat k - \hat q)
\end{eqnarray}
where $\Gamma_{s(t)}(\hat k , \hat q, \alpha)$ is the effective interaction in the singlet(triplet) channel.  
Figure \ref{diagrams} displays the Feynman diagrams which contribute to these leading terms in the 
perturbative expansion.  
In the second stage, the renormalization group flows of these effective interactions are computed and  
 the superconducting transition temperature is related to the energy scale at which these RG flows break down.  Following this basic prescription, one obtains
\begin{equation}
T_c \sim W e^{-1/\vert \lambda^{s,t}_0 \vert}
\end{equation}
where $\lambda_0^{s,t}$ are the most negative eigenvalue of 
\begin{eqnarray}
g_{s,t}(\hat{k}, \hat{q}, \alpha) = \sqrt{\frac{\bar{v}_f}{v_f(\hat k)}} \Gamma_{s,t}(\hat{k}, \hat{q}, \alpha)\sqrt{\frac{\bar{v}_f}{v_f(\hat q)}} 
\end{eqnarray}
The most negative eigenvalue determines the superconducting ground state and its eigenfunction is related to the pair wavefunction, as described in Ref. \cite{Raghu2010}.   
\begin{figure}
\begin{center}
\includegraphics[width=3.0in]{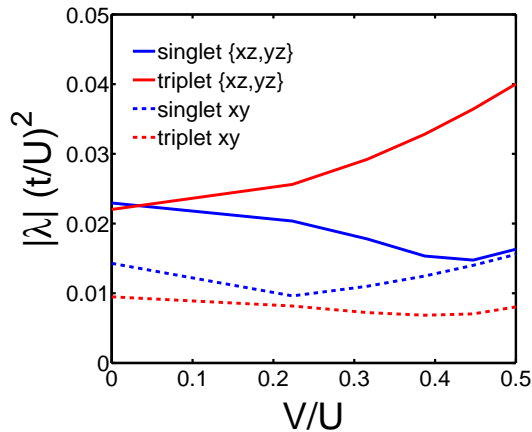}
\caption{Pairing eigenvalues as a function of $V/U$ for the bandstructure parameters 
quoted in the text.  The strongest pairing strengths occur among the quasi-1D $\{\alpha, \beta\}$ bands.  There is a near 
degeneracy of the singlet and triplet eigenvalues for $V =0$, but with  $V > 0$, the quasi-1D
triplet state is the dominant superconducting configuration. }
\label{lambda}
\end{center}
\end{figure} 

The central result of this section is shown in Fig. \ref{lambda}.  When $V=0$, the two dimensional 
xy band has its dominant pairing instability in the $d_{x^2-y^2}$ channel with a substantially lower pairing strength in the triplet channel.  However, the xz,yz bands exhibit a close competition between singlet and triplet pairing and therefore, triplet pairing is a close contender among these degrees of freedom.  
A similar competition between singlet and triplet 
pairng is thought to occur in the quasi-one dimensional organic Bechgaard salts where similar bandstructure effects are encountered\cite{Stuartbrown1,Bechgaard2012,Saito1987}.  

\begin{figure}
\begin{center}
\includegraphics[width=5.0in]{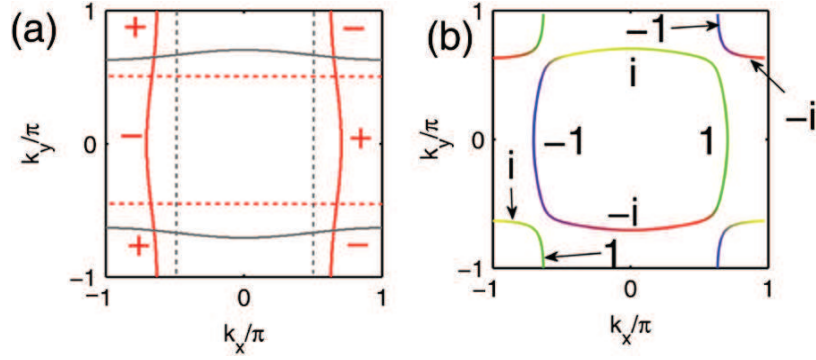}
\end{center}
\caption{(a) Triplet pair-wave function with sign structure shown for the d$_{xz}$ orbital when interorbital coupling is absent.  (b) Structure of the chiral p-wave superconductor when the interorbital mixing is taken into account.    }
\label{wf3}
\end{figure}

The triplet pair wave-function 
\begin{equation}
\Delta_{\alpha}(\vec k)  = i \left[ \vec d_{\alpha}(\vec k) \cdot \vec \sigma  \sigma^y \right], \alpha = xz,yz
\end{equation}
 is specified by the  complex vector $ \vec d(\vec k)$ in spin-space; in the absence of any mixing among the orbitals, its orientation is completely arbitrary.  Figure \ref{wf3}(a) shows the sign of $\Psi_x(\vec k) $ on the xz Fermi surface.  In addition to having odd parity, the wave function has two point nodes on the Fermi surface near $k_y = \pm \pi/2$ and is well approximated by  $ \vec d_x(\vec k) = \sin{k_x} \cos{k_y}\hat \Omega$ with $\hat \Omega$ an arbitrary unit vector in spin-space.  The nodal structure arises due to the peak in the susceptibility, and therefore the effective $\vec k$-dependent repulsive interaction at $\vec k = (2k_f^{1d}, \pi)$ which acts to enforce a sign change of the gap function on the Fermi surface.  By symmetry, the pair-wave function $\vec d_y(\vec k) = e^{i \phi} \sin{k_y} \cos{k_x} \hat \Omega'$ on the yz Fermi surface, and both $\phi$ and $\hat \Omega'$ are completely arbitrary  when the two orbitals are perfectly decoupled.  

\subsection{ Effect of $\delta H$ }
 The analysis above was carried through for $\delta H = 0$.   Next, we consider the effect of hybridization among the different orbitals and study the Hamiltonian 
 \begin{eqnarray}
 \delta H &=&  \sum_{\vec k \sigma}\left( c^{\dagger}_{\vec k , xz, \sigma} c^{\dagger}_{\vec k , yz, \sigma} \right) \left( \begin{array}{cc}
0 & g(\vec k) \\
g(\vec k) & 0 \end{array} \right) \left( \begin{array}{c} c_{\vec k , xz  \sigma} \\ c_{\vec k, yz, \sigma} \end{array} \right) \nonumber \\
&& + \lambda \sum_{\alpha, \beta} \sum_{\sigma \sigma'} \sum_{\vec k} c^{\dagger}_{ \vec k, \alpha,  \sigma} c_{\vec k , \beta, \sigma'}  \vec \ell_{\alpha \beta} \cdot \vec \sigma_{\sigma \sigma'}
\end{eqnarray}
where $g(\vec k) = -4t'' \sin{k_x}\sin{k_y}$, $t''=0.1t$, the second quantity above is the spin orbit coupling, and the angular momentum operators are expressed in terms of the totally anti-symmetric tensor as $\ell^a_{\alpha \beta} = i \epsilon_{a \alpha \beta}$.  Recent electronic structure calculations along with angle-resolved photoemission (ARPES) 
studies have produced the estimate $\lambda \approx 0.1t = 0.1 eV$\cite{Haverkort2008, Liu2008}.  
  
  For non-zero $\delta H$,  the relative phase and orientation 
  of $\vec d_x$ and $\vec d_y$ are no longer arbitrary.  A self-consistent calculation of the Bogoliubov-de Gennes (BdG) Hamiltonian taking into account $\delta H$ and the same functional form of the pair wave functions obtained in the previous section reveals that two distinct types of saddle point configurations are possible for the order parameter: (i) a configuration in which $\hat \Omega, \hat \Omega' \perp \hat z, \hat \Omega \cdot \hat \Omega' = 0, \phi = 0$, and (ii) $\hat \Omega =  \hat \Omega' = \hat z, \phi = \pm \pi/2$; all other configurations have a higher free energy and can therefore be neglected.  While the former saddle point preserves time-reversal symmetry with a non-collinear order parameter in spin-space, the latter saddle point breaks time-reversal symmetry, has a collinear order parameter, and corresponds 
  to the chiral p-wave state.  Although we have found that the chiral phase is energetically favored in the presence of spin-orbit coupling, the energy difference between these two possibilities is $\mathcal{O}(\lambda/t)^2$ and is far too small for us to make any robust predictions in our theory.  Therefore, we proceed with a phenomenological approach: we consider a free energy of the form  
  \begin{eqnarray}
  \label{f}
&& \mathcal{F} = \sum_{i =x,y}\left( a_z  \vert d^z_i \vert^2 + a_{\perp}  \vert \vec d^{\perp}_{i} \vert^2 \right) + u_1 \left( \vert \vec d_x 
\vert^2 + \vert \vec d_y \vert^2 \right)^2  \nonumber \\ 
&& + u_3 \left( \vec d_x \cdot \vec  d_y^* + \vec d_y  \cdot \vec d_x^* \right)^2 
+ u_4 \vert \vec d_x \times \vec d_y^* \vert^2 + u_2 \vert \vec d_x \vert^2 \vert \vec d_y \vert^2 \nonumber \\
\end{eqnarray}
where  we have neglected the gradient terms for sake of brevity.  Both
$a_{\perp}- a_z$ and $ u_3 - u_4 \sim \mathcal O(\lambda/t)^2$, and when $a_{\perp}> a_z$, the chiral state is favored but
only slightly more than the non-chiral solutions with $\vec d$ in plane.  
These small energy differences   imply that the d-vector is only 
weakly pinned along the c-axis, and is  corroborated by the 
absence of a change in the Knight shift for magnetic fields applied along the c-axis\cite{Murakawa2004} 
as well as for fields applied in the basal plane\cite{Ishida1998}.  In our subsequent analysis, we will assume that the chiral state has a lower free energy and will study an order parameter of the form
 \begin{equation}
\vec d = e^{i \theta} \left(  d_x + i  d_y  \right) \hat{z}, \ d_x, d_y \in \mathcal R
\end{equation}
where $\theta$ is the overall phase of the two condensates.  The order parameter structure on the Fermi surface is shown in Fig. \ref{wf3}(b).

Next, we address the issue of nodes on the Fermi surface when $\delta H \ne 0$.  We have studied   the  BdG Hamiltonian for the chiral state with the gap functions derived in the previous section.  The spectrum of such a Hamiltonian is readily obtained, and it is found that in order for the nodes to occur on the Fermi surface, the following conditions must be satisfied:  $g(\vec k) = 0$, and $\vec d_{x}(\vec k) = \vec d_y(\vec k) = 0$.  This in turn implies that even an infinitessimal  $g(\vec k)$  is sufficient to destroy the nodes.  Although the nodes are not topologically stable, it still follows that the gap becomes parametrically small on portions of the Fermi surface so long as $g(\vec k) \ll t $, and the gap minima are $\sim \Delta_0 \left( t''^2+\lambda^2\right)/t^2 $, where $\Delta_0$ is the maximum value of the gap on the Fermi surface.  The energy scale of these gap minima is therefore two orders of magnitude smaller 
than the transition temperature and their effects are expressed in the form of power laws in the specific 
electronic heat and nuclear spin relaxation rate.

\section{Criticism of the weak-coupling theory}
Our theory can be criticized on several possible grounds.  Firstly, it goes against the standard and intuitive notion that spin-triplet pairing arises when the dominant fluctuations are ferromagnetic in character.  Secondly, given that the $\gamma$ sheet is close to a van Hove point, the effects of electron correlations in general, and ferromagnetic fluctuations in particular, ought to be enhanced in this band.  Thus, by finding weak pairing strength in the $\gamma$ sheet, our theory ostensibly is missing some of the salient physics at play.  We will address each of these criticisms below.

It is easy to see why ferromagnetic fluctuations can enhance the tendency towards triplet pairing.  As an example, consider a system with kinetic energy $\epsilon(k) = k^2/2m$ for simplicity.  In such a system, the Fermi surface is a sphere centered around $k=0$.  In the presence of short-range repulsive interactions, $U>0$, the effective interaction in the triplet channel, shown in diagram (c) of Fig. \ref{diagrams} is
\begin{equation}
g_t(\hat k, \hat q) = -U^2 \chi(\hat k - \hat q) + \mathcal(O(U^4)) \simeq -U^2 \rho \left[ 1 + a \vert \hat k - \hat q \vert^2 + \cdots \right] + \mathcal(O(U^4)) 
\end{equation}
where we have made use of a long-wavelength expansion of the susceptibility, appropriate for systems with rotational invariance.  Here, $\rho$ is the density of states at the Fermi level, and $a \sim d^2 \chi/d q^2$ is a number that depends on microscopic details.  Since the pair wave-function has odd parity, the only component of $g_t$ which contributes to the triplet eigenvalue in the long wavelength limit is
\begin{equation}
g_t(\hat k, \hat q) =  V_1  \hat k \cdot \hat q   + \mathcal(O(U^2)) 
\end{equation}
where $V_1 = 2U^2  \rho a$.  This is a separable interaction in the triplet channel.  It leads to non-trivial solutions when $a < 0$, i.e., when the susceptibility is {\it peaked} at $q=0$, implying in turn that the dominant magnetic fluctuations are ferromagnetic.  

\begin{figure}
\begin{center}
\includegraphics[width=4.0in]{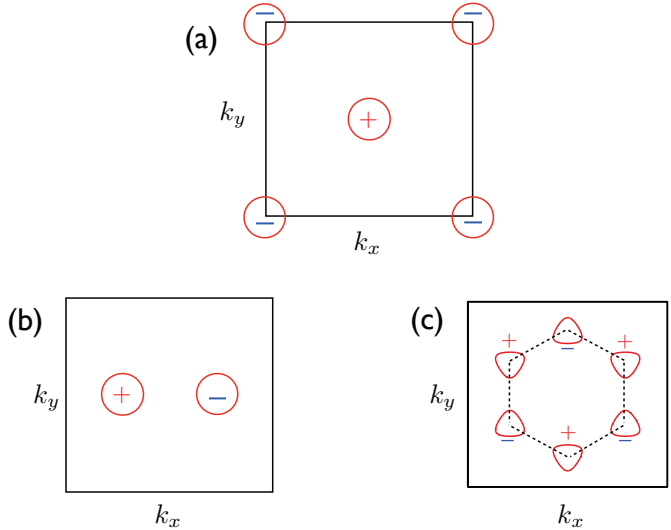}
\caption{A system with two Fermi pockets in which large momentum transfer particle-hole fluctuations give rise to pair wavefunctions that change sign from one pocket to the other.  (a) If the pockets were centered at $\bm k = (0,0)$, and $\bm k = (\pi, \pi)$, the superconductor would be a sign-changing s-wave (i.e. extended s-wave).  (b) If the pockets were centered at incommensurate momenta $\pm k_x$, the sign-changing state would have $p_x$-wave symmetry.  (c) In hexagonal systems, the same sign-changing solution would have f-wave symmetry.  Clearly all would be classified as having the same pairing ``mechanism" but only the first state is characterized by a spin singlet order parameter.  Ferromagnetic fluctuations are irrelevant here: yet, the latter two states exhibit spin triplet pairing.   }
\label{honeycomb}
\end{center}
\end{figure}

While this is a sensible argument, it does not imply a one-to-one correspondence between spin-triplet pairing and ferromagnetic fluctuations; at best, it is a helpful rule-of-thumb.  As a concrete counterexample, consider a system with two Fermi pockets, which has particle-hole fluctuations that are peaked at large momentum transfer.  If such a system were to exhibit superconductivity from repulsive interactions, the pair wavefunctions would change sign from one pocket to the other (see Fig. \ref{honeycomb}).  Depending on the precise location of the pockets and on the lattice point group symmetry, however, different pairing symmetries will result from exactly the same pairing mechanism.  For instance, on a square lattice, if the two Fermi pockets were centered at incommensurate points in the Brillouin zone, the system would exhibit p-wave pairing.   In a triangular, honeycomb, or Kagome lattices, when each pocket is centered at the hexagonal Brillouin zone corner(Fig. \ref{honeycomb}), the gap function has f-wave symmetry.  
Our conclusions  have  little to do with Ferromagnetic fluctuations.  

Other examples include the quasi-one dimensional organic superconductors which, due to peaks in the susceptibility at large momentum $\bm q = (2k_F, \pi)$, leads to sign changing singlet and triplet order parameters that have almost identical pairing strength.  While the singlet state has a higher pairing strength, application of a weak, in-plane magnetic field is believed to tilt the balance in favor of the spin-triplet state.  It makes little sense to say that the field induces ferromagnetic fluctuations; instead, it plays a role of a perturbation to change the condensation energy of the triplet state relative to the singlet state.

Next, we address the role of the $\gamma$ band being proximate to a van Hove singularity and the role of quasiparticle mass enhancement.  While mass enhancement does suggest that the effects of electron correlations are important, the fact remains that for $T_c < T \lesssim 40K$, the normal state is a well-behaved, albeit dressed Fermi liquid.  This in turn implies that the system is governed by a free fermion fixed point, and is destabilized only by the superconducting instability.   Thus, while the strong correlations lead to a heavy Fermi liquid, the quasiparticles themselves are coherent and weakly coupled; this means that the mass enters mainly in the form of  quasiparticle bandstructure parameters (i.e. hoppings) rather than as a bare onsite interaction, and a weak-coupling theory is nevertheless feasible.  This philosophy is consistent with the experimental fact that though the system is nearly nested, and though the $\gamma$ band is close to a van Hove singularity, neither antiferromagnetism nor ferromagnetism occur in the system.

\section{Phenomenological consequences of the weak-coupling theory}

Throughout this article, we have stressed microscopic aspects of superconductivity in Sr$_2$RuO$_4$.  However, it is important to mention that there are several phenomenological consequences of our theory that have some support from experimental observations.  Firstly, as was shown in Refs. \cite{Kallin1,Kallin2,Annett1}, there is an {\it intrinsic}  Kerr response in a multi-band chiral superconductor, whereas in the single-band case, translational invariance guarantees that at least to leading order, there is zero intrinsic Kerr response (vertex corrections, which are suppressed by powers of $v_f/c$, provide a finite but small correction).  The authors of Ref. \cite{Kallin1, Kallin2} found that by invoking a multi-band superconductor, the resulting Kerr angle is comparable to that found in experiments.  For a particularly sharp prediction involving a nearly {\it resonant} Kerr response, we urge the reader to consult Ref. \cite{Kallin2}.  It would be interesting to see whether indeed such a resonance can be found when the frequency of the electromagnetic radiation is tuned to the scale characteristic of the weak inter-orbital couplings.  

In the remainder of this section, we will discuss two additional phenomenological consequences of our theory.  Firstly, we will discuss edge modes and their implication for edge currents.  The goal is to elaborate on the discussion presented in Ref. \cite{Raghu2010a}.  Secondly, we will discuss the possibility of sharp, low energy collective excitations that occur in our theory, which are discussed in more detail in Ref. \cite{Chung2012}.  

\subsection{ Topological properties and edge currents }

We consider the topological properties of the quasi-1D superconductor assuming that $a_z < a_{\perp}$ in Eq. \ref{f}  so that the chiral state is the ground state.  In this case, the z-component of the spin of the condensate remains conserved, and acts as a  $\mathcal U(1)$ `charge' degree of freedom.  
Thus, we can define the Nambu spinor in momentum space using the band indices as $\Psi_{\nu \vec k} = \left( d_{ \vec k, \nu, \uparrow}, d^{\dagger}_{- \vec k, \nu, \downarrow} \right)^T$, $\nu = \alpha, \beta$, the BdG Hamiltonian in this basis is described in terms of two Anderson pseudospins, one for each Fermi surface: 
\begin{equation}
H_{BdG} = \sum_{\nu = \alpha, \beta} \Psi^{\dagger}_{\nu \vec k }\left[ \vec \delta_{\nu}(\vec k) \cdot \vec \tau_{\nu} \right] \Psi_{\nu \vec  k }.
\end{equation}
For a chiral p-wave state of the form $\Delta_{\nu} \left( s_x(\vec k), s_y(\vec k) \right)$ on the $\nu$-th Fermi surface, where $s_x(s_y)$ are functions that transform under point group operations as $k_x(k_y)$, we have the following explicit form for the Anderson pseudospin operator: 
\begin{equation}
\delta_{\nu}(\vec k) = \left( \Delta_{\nu} s_x(\vec k), \Delta_{\nu} s_y(\vec k), \epsilon_{\nu \vec k} - \mu \right)
\end{equation}
For the chiral state, the pseudospin has the form of a skyrmion in momentum space: it points along the $-\hat z(+\hat z)$-direction inside(outside) the Fermi surface, and on the Fermi surface, it lies in plane, winding by $2 \pi$ around the Fermi surface.  The topological properties of the chiral state comes from the integer skyrmion number which is defined as 
\begin{equation} 
\mathcal N_{\nu} = \frac{1}{4 \pi} \int d^2 k  \hat  \delta_{\nu} \cdot \left( \partial_x \hat \delta_{\nu} \times \partial_y \hat \delta_{\nu}\right)
\end{equation}
where $\hat \delta_{\nu} = \vec \delta_{\nu}/ \vert \vec \delta_{\nu} \vert$.  The quantity $\mathcal N_{\nu}$ is insensitive to smooth perturbations of the system, provided that the gap to quasiparticle excitations remains non-zero.  Physically, the net number of chiral (i.e. the number of left-moving minus the number of right-moving) quasiparticle modes at the edge of the superconductor is given by the skyrmion number, and so long as $\mathcal N_{\nu} \ne 0$,they cannot be localized by backscattering.

The sign of $\mathcal N_{\nu}$ can change only when the skyrmion is altered to an anti-skyrmion.  This can be accomplished by flipping the sign  of an {\it odd} number of components of $\hat \delta$.  Up to overall gauge transformations, there are two distinct ways of changing the skyrmion into an antiskyrmion: (i) a time-reversal operation under which the chirality of the order parameter is changed from $p_x + i p_y \rightarrow p_x - i p_y$, or (ii) the transformation $\epsilon_{\nu} - \mu \rightarrow \mu - \epsilon_{\nu}$ which changes the Fermi surface dispersion from electron-like to hole-like or {\it vice versa}.  As a corollary of the latter scenario, it follows that when the Fermi surface consists of both an electron and a hole pocket - as is the case for the quasi-1D $\alpha, \beta$ surfaces - a $p_x + ip_y$ superconductor has a net skyrmion number of {\it zero}: i.e. it is not a topological superconductor.    

As a consequence, the chiral edge modes along the boundary of the superconductor and along domain 
walls are no longer protected: in the presence of disorder, these modes are localized due to the existence of backscattering channels.  We can understand this more explicitly by considering a semi-infinite superconductor for $x< 0$, letting $x>0$ be the vacuum.  The quasiparticle edge modes are zero energy solutions of $H_{BdG}$; excellent approximations  for for their wave functions can be obtained by linearizing the dispersion about each Fermi surface:
\begin{equation}
\Psi^0_{ \nu}(x) \propto  e^{ - i {\rm sgn}(v^{\nu}_f) k_y y } e^{ - \int_0^x dx' \vert \Delta_{\nu}(x') \vert /\vert v_f^{\nu} \vert } \chi_{\nu}
\end{equation}
where we have defined the electron(hole) pocket to have positive(negative) Fermi velocity, and $\chi$ is a spinor which is an eigenstate of $\tau_{\nu}^y$.   Thus, in 
the quasi-1D superconductor, there are two sets of quasiparticle modes, one from each Fermi pocket, which counterpropagate along the boundaries (in this case along the y direction) and along domain walls of the system.  Although their propagation velocities are different, in the presence of disorder, localization will still result.  

Having discussed the neutral quasiparticle edge modes, we next consider the electrical currents in a simplest chiral 
superconductor having a single Fermi surface.  Due to the breaking of time-reversal symmetry, there are two contributions to these edge currents.  Firstly, there is a bulk contribution which originates from the multi-component nature 
of the order parameter.  In the Ginzburg Landau approximation, the contribution arises from  a gradient term of the form
\begin{equation}
J^a_{bulk} = -i \epsilon^{ab} K \left( \vec d_x \partial_b \vec d^*_y - c.c \right)
\end{equation} 
with $K \sim \mathcal{O}(\Delta/E_f)^2$.  In the continuum limit, it can be shown that  this term is proportional to the curl of the total angular momentum density of the condensate.  The second contribution to the electrical current comes from the 
chiral quasiparticle modes near the edge of the system.  In a single-band system, there is no reason for these currents to be suppressed.  Moreover, there are sharp predictions of mesoscopic effects in such systems, including the possibility of field-induced reentrant superconductivity\cite{Huo2012}.  The fact that edge currents have not been observed in highly sensitive scanning squid probes\cite{bjornsson2005,Kirtley2007, Hicks2010} leads us to question further the simple one-band descriptions that have prevailed in the field.  By contrast, in our quasi-one dimensional superconductor, the magnitude of  the electrical current is determined by the small interorbital couplings, since the chiral nature of the condensate is established only by these terms.  Since the $p_x$ and $p_y$ components of the condensate are nearly decoupled, (i.e. because interorbital couplings play an important role only in regions of $\bm k-$space where the Fermi surfaces nearly cross), we can reasonably expect that there will be a substantial reduction of edge currents in our theory as compared to theories based on a single-band description of the chiral state.      If the $\{\alpha, \beta\}$ bands were the active bands, at lower temperature when substantial pairing strength develops on the $\gamma$ band, the expectation remains that observable edge currents ought to be present in the system.  
 A more quantitative theory of edge currents  shall be presented 
elsewhere\cite{Raghuinprep}.  

\subsection{Collective modes}

A superconductor described by a multi-component order parameter ($p_x$ and $p_y$ in the present context), will have collective mode excitations associated with the relative phase difference, $\phi_-\equiv\theta_{x}-\theta_{y}$, where $\theta_x(\theta_y)$ is the phase of the $p_x(p_y)$ component of the order parameter.
At zero temperature, such a mode would be expected to have a frequency $\hbar\omega_0\sim \sqrt{{\cal J} /\chi}$, where $\chi \sim N(0)$ is the compressibility ($N(0)$ is the density of states at the Fermi energy) and ${\cal J}$ is the second derivative of the condensation energy with respect to $\phi_-$. Given that the condensation energy $\sim N(0) |\Delta_0|^2$ where $|\Delta_0|$ is the root mean squared gap magnitude, this 
means that $\hbar\omega_0 \sim |\Delta_0|$ for a single-band superconductor.  Thus, in a single-band chiral superconductor, the characteristic energy scale of the collective modes is of the order of the gap, meaning that the modes are likely to be damped by the continuum.  Similar considerations apply to fluctuations in the relative orientation of the spins ({\it i.e.} the $d$-vector) in a two-component triplet superconductor.

By contrast, %in the limit in which
if the two components of the order parameter %$p_x$ and $p_y$,
are associated with different orbitals, {\it i.e.} the $p_x$ component with the $d_{xz}$ and the $p_y$ with $d_{yz}$ orbital respectively, and if mixing between the different orbitals were absent, then collective fluctuations of the relative phase and spin-orientation would be gapless.
The orbital-mixing terms, denoted as $\delta H$ above, result in a non-vanishing dependence of the condensation energy on $\phi_-$.
As a result, if the superconductivity arises primarily from the quasi-1D bands, then the relative phase mode is expected to have an energy $\hbar\omega_0 \approx \gamma \Delta_0$, where $\gamma$ vanishes continuously as $\delta H$ tends to zero. 
Naturally, similar considerations apply to the relative spin orientational fluctuations. 
The relevant question here is whether we should have $\gamma \sim 1$ or $\gamma \ll 1$ in the physical limit of $\vert \Delta_0 \vert \ll \delta H \ll W$, where $W$ is the bandwidth. The recent work by Chung {\it et al.}\cite{Chung2012} showed that $\gamma$ will scale with $\delta H/W$, with almost no dependence on $\Delta_0$. This is because ${\cal J}$ is the cost in the pair condensation energy due to fluctuations of the relative phase  and spin orientation, and fluctuations of these quantities will affect the pair condensation energy mainly in the portion of the Fermi surface that is affected by the orbital hybridization, the size of which is determined almost entirely by $\delta H/W$.  Thus, if superconductivity arises from the quasi 1D bands, the collective modes should have low energies for the same reason that the edge currents are small.

\section{Conclusion}

We have stressed that the microscopic origin of superconductivity in Sr$_2$RuO$_4$ remains unknown.  Most workers in the field believe that the physics of Sr$_2$RuO$_4$ is analogous to superfluid $^3$He, where dominantly ferromagnetic fluctuations of a nearly isotropic Fermi surface produce triplet pairing.  By contrast, we have found that the material better resembles the organics, Fe-based and cuprate superconductors in the sense that particle-hole fluctuations with large momentum transfer  are responsible for spin triplet pairing in this system.    From the experimental standpoint, the heat capacity measurements that have often been cited as evidence in favor of $\gamma$ being the active band, do not rule out the complementary picture in which $\{ \alpha, \beta \}$ are the active bands.  In a well-controlled weak-coupling treatment, appropriate for the description of an unconventional superconductor that condenses from a Fermi liquid, $\{\alpha,\beta \}$ are the active bands.  
 The resulting superconducting state has a `hidden' quasi-one dimensionality: although it preserves the tetragonal symmetry of the underlying lattice, it consists of two nearly decoupled quasi-one dimensional condensates.  In addition, the solution obtained in this limit has some phenomenological support, in  that as a multi-band system, it allows for an intrinsic Kerr response\cite{Kallin1, Kallin2, Annett1}.    Moreover, since the $p_x$ and $p_y$ components are nearly decoupled, ``living" on different orbitals,  the electrical currents along boundaries of this system can be substantially reduced.   Another important consequence of the fact that the $p_x, p_y$ components are nearly decoupled is that there are sharp, low energy collective modes that correspond to inter-orbital relative phase and relative spin-orientation fluctuations.  These modes are ``almost-Goldstone modes" and have energy well below the pairing gap.  It will be vital to test some of these phenomenological predictions in the near future. It is our hope that the ideas described here will lead to further developments in the field, both along experimental and theoretical fronts.   
  
{\bf Acknowledgments}
We thank D. Agterberg, J. Alicea, E. Berg, S. Brown, T. Devereaux, C. Hicks, C. Honerkamp, C. Kallin, A. Kapitulnik, J. Kirtley, S. Kivelson, A. P. Mackenzie, K. Moler, D. Scalapino,  R. Thomale, and F.-C. Zhang for helpful discussions.  This work is supported  by the Department of Energy, Office of Basic Energy Sciences, Division of Materials Sciences and Engineering, under contract DE-AC02-76SF00515, and the Alfred P. Sloan Foundation (SR).

\section*{References}

\bibliography{SrRuO214_m2s}

\end{document}